\documentclass[pre,preprint,showpacs,amsmath]{revtex4}

\usepackage{graphicx}
\usepackage{bm}

\renewcommand{\vec}[1]{\hm{#1}}
\newcommand{\cn}{\mbox{cn}}

\begin{document}

\title{Breathers or quasibreathers?}
\author{G.M. Chechin}
  \email{chechin@phys.rsu.ru}
\author{G.S.Dzhelauhova}
\author{E.A.Mehonoshina}
\affiliation{Department of Physics, Rostov State University, Russia}
\date{\today}
\begin{abstract}
For the James breathers in the $K_2-K_3-K_4$ chain and for breathers
in the $K_4$-chain, we prove numerically that these dynamical
objects are not strictly time-periodic. Indeed, for the both cases,
there exist certain deviations in the vibrational frequencies of the
individual particles, which certainly exceed the possible numerical
errors. We refer to the dynamical objects with such properties as
quasibreathers. For the $K_4$-chain, a rigorous investigation of
existence and stability of the breathers and quasibreathers is
presented. In particular, it is proved that they are stable up to a
certain strength of the intersite part of the potential with respect
to its on-site part. We conjecture that the main results of this
paper are also valid in the general case and, therefore, it seems
that one must speak about quasibreathers rather than about strictly
time-periodic breathers.
\end{abstract}
\maketitle

\section{Introduction}
According to the conventional definition
\cite{Aubry,flach-1,Flach-2}, discrete breathers are spatially
localized and time-periodic excitations in nonlinear lattices.
Because of the space localization, different particles vibrate with
essentially different amplitudes. On the other hand, it is typical
for nonlinear systems that frequencies depend on amplitudes of
vibrating particles. Therefore, it is not obvious how a discrete
breather can exist as an exact time-periodic dynamical object
because, in this case, the particles with considerably different
amplitudes must vibrate with the same frequency. Surprisingly, we
did not find an explicit answer to this question in the literature
on discrete breathers.

This paper is devoted to some aspects of the above problem. In
Sec.~\ref{jambreath}, we consider the breathes introduced by James
in Ref.~\cite{l4} and arrive at the conclusion that there are some
deviations of the vibrational frequencies of the individual
particles from the average breather frequency and these deviations
certainly exceed the possible numerical errors.

On the other hand, the analytical form of the discrete breathers
used in \cite{l4,l5} is not an exact solution to the nonlinear
dynamical equations of the $\ K_{2}-K_{3}-K_{4}$ chain and,
therefore, one can suspect that the above deviations are induced by
an inaccuracy of the initial conditions for solving the appropriate
Cauchy problem.

To establish results beyond suspicion, we consider discrete
breathers in a nonlinear chain with a {\it uniform} on-site and
intersite potential of the forth order (see Sec.~\ref{k4}). In other
words, we study the $\ K_{2}-K_{3}-K_{4}$  chain with $K_2=K_3=0$
and call it "$K_4$-chain". For this case, there exists a localized
nonlinear normal mode (NNM) by Rosenberg \cite{l6, l7} which
represents an exact discrete breather (DB). Let us note that DBs for
such potentials were discussed in a number of papers (see, for
example, Refs.~\cite{l8, l9}), but from a somewhat different point
of view. Here, we obtain practically exact form of the DB and study
its stability. It turns out that {\it any infinitesimal vicinity} of
the exact breather solution consists of {\it stable} dynamical
objects (for the appropriate strength of the intersite potential)
which {\it are not} time-periodic. The strict periodicity occurs
only along a certain line in the space of possible initial
conditions which give rise to NNM. All other initial conditions
generate the dynamical objects which can be considered as {\it
quasibreathers}, because they correspond to  quasiperiodic motion.
As a consequence, there are some deviations in the vibrational
frequencies of the individual breather's particles similar to those
for the James breathers.

Since, in every physical or computational experiment, we cannot tune
exactly onto a line in the many-dimensional space of all possible
initial shapes of the desired periodic solution, it is reasonable to
speak only about quasibreathers. It seems that such situation occurs
not only for the considered case admitting the exact solution, but
also for the general case.

In connection with the above mentioned term "quasibreathers", let us
note that the term "quasiperiodic breathers" is used in literature
for different dynamical objects (see Conclusion to this paper).

\section{James breathers \label{jambreath}}
An approximate analytical form of breathers with small amplitudes
for the $\ K_{2}-K_{3}-K_{4}$  chain was obtained in \cite{l4}. Some
computational experiments with these breathers were presented in
\cite{l5}. The main results of \cite{l4, l5} can be outlined as
follows.

Let us consider a nonlinear chain of $\tilde{N}=2N+1$ identical
masses ($m=1$) which are equidistant in the equilibrium state.
Interaction only between the nearest neighboring particles is
assumed. Then dynamical equation for the $\ K_{2}-K_{3}-K_{4}$ chain
reads
\begin{equation} \label{eq1}
\ddot{x}_{n}=V'(x_{n+1}-x_{n})-V'(x_{n}-x_{n-1}), \;\;\;\; n=-N..N,
\end{equation}
\begin{equation} \label{eq2}
V(u)=K_{2}u^{2}+K_{3}u^{3}+K_{4}u^{4}.
\end{equation}
Here $x_{n}$ is a displacement of $n$-th particle from its
equilibrium position, $V(u)$ and $V'(u)$ are the potential of the
interparticle interaction and its derivative, respectively.

It was proved by James that for any $\omega_{b} $ (breather
frequency) slightly exceeding the maximal phonon frequency
$\omega_{max}$ (in our case, $K_2=\frac{1}{2}$ and, therefore,
$\omega_{max}=2$), i.e. $\omega_{b}^2=4+\mu,\;\;\;\mu\ll1$ , there
exists the following breather solution to
Eqs.~(\ref{eq1},\ref{eq2}):
\begin{equation}\label{eq3}
\ y_{n}(t) =
(-1)^{n}\sqrt{\frac{2\mu}{B}}\frac{\cos(\omega_{b}t)}{\cosh\left(n\sqrt\mu\right)},
\end{equation}
where $B=\frac{1}{2}V^{(4)}(0)-(V^{(3)}(0))^2$,
\begin{equation}\label{eq4} y_n\equiv
V'(u_n)=2K_2u_n+3K_3u_n^2+4K_4u_n^3, \;\;\;\;u_n=x_n-x_{n-1}.
\end{equation}
Here $y_n$ are new variables introduced instead of the old variables
$x_n$ (actually these new variables represent forces acting on the
particles of the chain).

Thus, Eq.~(\ref{eq3}) determines a {\it family} of breather
solutions. Indeed, there exist a breather with amplitude
proportional to $\sqrt\mu$ for any fixed frequency $\omega_b$. The
smaller the deviation of the breather frequency $\omega_b$ from the
maximal phonon frequency 2, the less the amplitude of the breather.
For the case $\mu\rightarrow0$, the hyperbolic cosine in the
denominator of (\ref{eq3}) goes to unity for all numbers $n$ and,
therefore, the breather localization get worse. Actually, in this
limit, breather tends to the extended $\pi$-mode with the
infinitesimal amplitude \footnote{\ In the $\pi$-mode, all the
particles vibrate with the same amplitude, while all the neighboring
particles are out-of-phase.}.

\begin{figure}
\includegraphics[width=135mm,height=75mm]{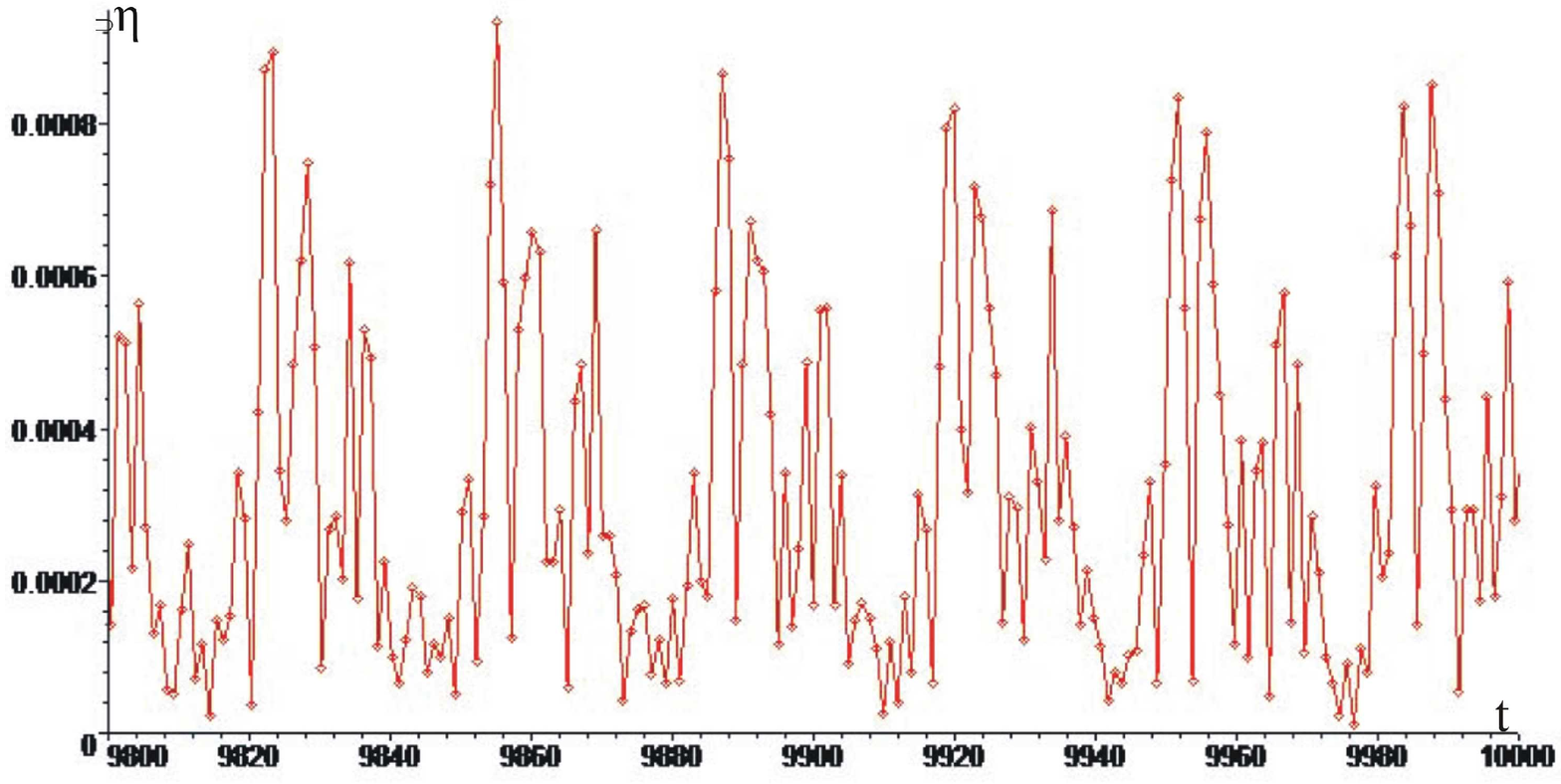}
\caption{\label{fig1} Mean square deviation $\eta(t)$ for the James
breather for long-time interval. All parameters are the same as in
Table \ref{t1}. Time $t$ is given in periods $T_0=\pi$ of the
$\pi$-mode.}
\end{figure}

Computational experiments reported in \cite{l5} have confirmed the
theoretical breather shape (\ref{eq3}). We tried to reproduce some
results of that paper, for example, those depicted in
Fig.~\ref{fig1}. To this end we started with Eq.~(\ref{eq3}) for
$t=0$, solved the cubic equations (\ref{eq4}) for obtaining the
initial conditions for $x_n(t)$, and then integrated numerically the
differential equations~(\ref{eq1}) of the $\ K_{2}-K_{3}-K_{4}$
chain.

Using the numerical values of $K_2$, $K_3$, $K_4$, $\omega_b$ and
other parameters from the paper \cite{l5}, we indeed obtained some
localized dynamical objects which seemed to be time-periodic at
first sight. But a closer examination revealed more complexity.

Bearing in mind the question posed in Introduction, we began to
follow the evolution of frequencies of the {\it individual}
particles participating in the breather vibration. Some results of
this analysis are presented below.

In Table \ref{t1}, we give the frequencies of nine particles
($n=-4..4$) near the center ($n=0$) of the breather which were
calculated within certain time intervals close to the instants $t_k$
listed in the first row of the table. Note that all these
frequencies $\omega_j(t_k)$ are sufficiently close to the breather
frequency $\omega_b=2.01$, which was used in Eq.~(\ref{eq3}), but
their {\it deviations} from $\omega_b$ certainly exceed the possible
numerical errors.

Let us comment on the computational procedure. We used the
fourth-order Runge-Kutta method with time step $h$ of about $0.0003
\ T_0$, where $T_0=\dfrac{2\pi}{\omega_{max}}=\pi.$ For the times
given in Table~\ref{t1}, our simulations conserved the total energy
of the chain up to $10^{-10}$. The frequencies of the individual
particles were obtained by calculating adjacent zeros of the
functions $x_j(t)$ in certain intervals near fixed instants $t_k$.
In turn, these zeros were computed by dichotomy and by Newton-Rafson
method.

It is expedient to introduce certain mean values characterizing
frequency deviations of the  individual breather particles. We
specify the mean value $\bar{\omega}(t_k)$ and the mean square
deviation $\eta(t_k)$ of different $\omega_j(t_k)$ for the breather
particles at the moment $t_k$ as follows:
\begin{equation}\label{eq10}
\bar{\omega}(t_k)=\frac{1}{M}\cdot\sum_{j}\omega_j(t_k),
\end{equation}
\begin{equation}\label{eq11}
\eta(t_k)=\sqrt{\frac{\sum_{j}[\omega_j(t_k)-\bar{\omega}(t_k)]^2}{M(M-1)}}.
\end{equation}
Here $j=-M..M\;\;(M<N)$ are numbers of the breather particles with
significant values of $x_j(t)$. The values $\bar{\omega}(t_k)$ and
$\eta(t_k)$  are given, respectively, in the two last rows of
Table~1. In the last column of this table, we give
\begin{equation}\label{eq12}
\sigma_j=\sqrt{\frac{\sum_{t_k}[\omega_j(t_k)-\bar{\omega}(t_k)]^2}{\tilde{M}(\tilde{M}-1)}}
\end{equation}
which represents the mean square deviation of the frequency for each
breather particle after averaging upon different moments (for this
averaging, we have used all the frequencies $\omega_j(t)$ which were
calculated up  to $t=t_k$ indicated it Table 1). [Note that $M$ in
Eqs.~(\ref{eq10},\ref{eq11}) is the number of considered breather
particles, while $\tilde{M}$ in Eq.~(\ref{eq12}) is the number of
$\omega_j(t)$ which were calculated].

\begin{table}
  \centering
  \caption{\label{t1} Deviations in frequencies of the individual particles
  for the James breather in $K_2-K_3-K_4$ chain with $\tilde{N}=101$ particles.
  Here $K_2=0.5$, $K_3=-0.1$, $K_4=0.25$, $\omega_b=2.01$, $\mu=0.0401$, $T_0=\pi$, $h=0.001$}
\begin{tabular}{|c|c|c|c|c|c|}
  \hline
    & $t_1=100\,T_0$ & $t_2=600\,T_0$ & $t_3=1000\,T_0$ & $t_4=1500\,T_0$ & $ \sigma $ \\
  \hline
  $\omega_{-4}$ & 2.0101459926 &  2.0097251337 &  2.0109780929 &2.0085043104 & 2.3348301075e-5\\
  $\omega_{-3}$ & 2.0068349042 &  2.0093263880 &  2.0070499430 &2.0100257555 & 2.4686634177e-5\\
  $\omega_{-2}$ & 2.0112977946 &  2.0089800520 &  2.0110369763 &2.0080931884 & 2.6957224898e-5\\
  $\omega_{-1}$ & 2.0065877923 &  2.0099366007 &  2.0070857334 &2.0101152322 & 2.9130307350e-5\\
  $\omega_0$    & 2.0117517055 &  2.0086968158 &  2.0110115290 &2.0080038280 & 2.9739839019e-5\\
  $\omega_{1}$  & 2.0065877923 &  2.0099366007 &  2.0070857334 &2.0101152322 & 2.9130307350e-5\\
  $\omega_2$    & 2.0112977946 &  2.0089800520 &  2.0110369763 &2.0080931884 & 2.6957224898e-5\\
  $\omega_3$    & 2.0068349042 &  2.0093263880 &  2.0070499430 &2.0100257555 & 2.4686634177e-5\\
  $\omega_4$    & 2.0101459926 &  2.0097251337 &  2.0109780929 &2.0085043104 & 2.3348301075e-5\\
  \hline
  $\bar{\omega}$  & 2.0090538525 &  2.009403685 &  2.00925700227051 & 2.0090534223 & \\
\hline
  $\eta$ &  7.61439353253e-4 &  1.5116218526e-4 & 6.9232221345e-4 & 3.2690761972e-4 &  \\
\hline
\end{tabular}
\end{table}

Some questions arise in connection with the result.
\begin{enumerate}
\item Because $\eta(t_k)$ are rather small, one can
suspect that they brought about by certain numerical errors. Is it
true?
\item Formula~(\ref{eq3}) represents only an approximation to the unknown exact
breather solution since it was obtained by neglecting some
higher-order terms \cite{l4, l5}. To what extent the appearance of
nonzero $\eta(t_k)$ reflects the properties of the real breather?
\item Is there any growth of $\eta(t_k)$ for large times? It is an important question because such growth, if exists, possibly
means the onset of stability loss of the exact breather solution.
\end{enumerate}

To shed some light on numerical errors problem, we computed
$\eta(t_k)$ and $\sigma_j$ for the $\pi$-mode which represents a
{\it strictly periodic} dynamical regime in any nonlinear chain
(see, for example, \cite{l10,l11}). Using the {\it same}
computational procedure as that for obtaining Table \ref{t1}, we get
Table \ref{t2} for the case of the $\pi$-mode vibrations. From the
latter table, one can see that deviations $\eta(t_k)$ and $\sigma_j$
for the $\pi$-mode \footnote{\ Note that the frequency of the
$\pi$-mode, in our case, is larger than $\omega_{max}$ of the phonon
spectrum ($\pi$-mode is an example of nonlinear normal modes in
anharmonic lattices).} turn out to be zero (up to machine
precision). Comparing these results with those from Table \ref{t1},
we conclude that deviations in vibrational frequencies of the
individual particles for the James breather {\it are not} numerical
errors.
\begin{table}
  \centering
  \caption{\label{t2} Frequencies of the individual particles for the $\pi$-mode with amplitude $A=0.1$. Here
  $K_2=0.5$, $K_3=-0.1$, $K_4=0.25$,  $T_0=\pi$, $h=0.001$}
\begin{tabular}{|c|c|c|c|c|c|}
  \hline
    & $100\,T_0$ & $600\,T_0$ & $1000\,T_0$ & $1500\,T_0$ & $\sigma$ \\
  \hline
  $\omega_{-4}$ & 2.00748367227 & 2.00748367227 & 2.00748367227 & 2.00748367227 & 0 \\
  $\omega_{-3}$ & 2.00748367227 & 2.00748367227 & 2.00748367227 & 2.00748367227 & 0 \\
  $\omega_{-2}$ & 2.00748367227 & 2.00748367227 & 2.00748367227 & 2.00748367227 & 0 \\
  $\omega_{-1}$ & 2.00748367227 & 2.00748367227 & 2.00748367227 & 2.00748367227 & 0 \\
  $\omega_0$ & 2.00748367227 & 2.00748367227 & 2.00748367227 & 2.00748367227 & 0 \\
  $\omega_1$ & 2.00748367227 & 2.00748367227 & 2.00748367227 & 2.00748367227 & 0 \\
  $\omega_2$ & 2.00748367227 & 2.00748367227 & 2.00748367227 & 2.00748367227 & 0 \\
  $\omega_3$ & 2.00748367227 & 2.00748367227 & 2.00748367227 & 2.00748367227 & 0 \\
  $\omega_4$ & 2.00748367227 & 2.00748367227 & 2.00748367227 & 2.00748367227 & 0 \\
  \hline
  $\bar{\omega}$ & 2.00748367227 & 2.00748367227 & 2.00748367227 & 2.00748367227 & \\
\hline
  $\eta$ & 0 & 0 & 0 & 0 & \\
\hline
\end{tabular}
\end{table}

The function $\eta(t)$ is depicted for large time intervals in
Fig.~\ref{fig1}. This function was found by calculating {\it all}
zeros of each displacement $x_j(t)$ and averaging the obtained
frequencies $\omega_j(t)$ with the aid of Eq.~(\ref{eq11}). From
this figure, it is obvious that $\eta(t)$ does not increase in
magnitude, and demonstrates certain oscillations similar to chaotic.

The above discussed behavior of the breather particles  will be
interpreted in the  next sections with an example of a model which
admits an exact breather solution.

\section{Existence of breathers in the $K_4$-chain \label{k4}}

We consider $2N+1$ particles chains with fourth-order potential and
periodic boundary conditions. The potential includes both on-site
and intersite parts. In contrast to the on-site terms, the intensity
of the intersite terms will be varied. We write this potential in
the form
\begin{equation}\label{eq8a}
U=\frac{1}{4}\sum_n x_n^4+\frac{\beta}{4}\sum_n (x_n-x_{n-1})^4.
\end{equation}

The Newton dynamical equations for such chain can be written as
follows:
\begin{equation}\label{eq30}
\ddot{x}_n=-x_n^3+\beta[(x_{n+1}-x_n)^3-(x_n-x_{n-1})^3],\;\;\;\
n=-N..N,
\end{equation}
\begin{equation}\label{eq31}
x_{N+1}(t)=x_{-N}(t),\;\;\ x_{-N-1}(t)=x_{N}(t).
\end{equation}
Here $\beta$ is the parameter characterizing the intensity of the
intersite potential. It will be shown that {\it stability} of the
breather solution in the chain (\ref{eq30}) depends essentially on
the value of $\beta$ (see below).

Models similar to (\ref{eq30}) have been considered in the papers
\cite{l8,l9}, but we analyze the chain (\ref{eq30}) with different
purposes and in a different manner.

It is well-known that space and time variables can be separated in
Eq.~(\ref{eq30}) and this was done in \cite{l8} (for the case without
on-site potential). We prefer to treat breather solution to
(\ref{eq30}) in terms of the nonlinear normal modes (NNM) introduced
by Rosenberg in \cite{l6,l7}. Indeed, it was proved that for any
{\it uniform} potential there exist (localized or/and delocalized)
NNMs which represent strictly periodic motion of all the particles
of the considered mechanical system. More precisely, equations of
motion of a many particle system for the dynamical regime
corresponding to a fixed NNM reduce to only one differential
equation for the displacement $x_0(t)$ of an arbitrary chosen
particle (this is the so called "leading" or "governing" equation),
while displacements $x_j(t)$ $(j\neq0)$ of all the other particles
are proportional to $x_0(t)$ at any instant $t$. Such dynamical
behaviour is remeniscent of the linear normal modes whose time
dependence is represented by sinusoidal functions because the
leading equation, in this case, is the equation of the harmonic
oscillator.

It is worth to mention that "bushes of modes" introduced in
\cite{l12} and investigated in a number of other papers
\cite{l13,l14,l15,l10,l11} represent a quasiperiodic motion because
we have $m$ leading differential equations for the $m$-dimensional
bush and, therefore, NNMs should be thought of as one-dimensional
bushes.

Below, we consider the procedure for obtaining NNMs. Assuming
\begin{equation}\label{eq32}
x_n(t)=k_nx_0(t),\;\;\;\ n=-N..N,
\end{equation}
for any time $t$ with constant coefficients $k_n$ and substituting
this expression into Eq.~(\ref{eq30}), we obtain
\begin{equation}\label{eq33}
k_n\ddot{x}_0=\{-k_n^3+\beta[(k_{n+1}-k_n)^3-(k_n-k_{n-1})^3]\}x_0^3,
\end{equation}
where \begin{equation}\label{eq34} k_{N+1}=k_{-N},\;\;\;
k_{-N-1}=k_N
\end{equation}
in accordance with the boundary conditions (\ref{eq31}).

The leading equation (it corresponds to $n=0$) reads:
\begin{equation}\label{eq35}
\ddot{x}_0=\{-1+\beta[(k_1-1)^3-(1-k_{-1})^3]\}x_0^3,
\end{equation}
because we can assume $k_0=1$.

Demanding all other equations (\ref{eq33}) to be {\it identical} to
Eq.~(\ref{eq35}), we obtain the following relations between the
unknown coefficients $k_n$  ($n=-N..N, \ n\neq0$):
\begin{equation}\label{eq36}
-k_n^3+\beta[(k_{n+1}-k_n)^3-(k_n-k_{n-1})^3]=k_n\{-1+\beta[(k_1-1)^3-(1-k_{-1})^3]\}.
\end{equation}

Thus, we arrive at the system of $2N$ algebraic equations with
respect to $2N$ unknowns $k_n$ (Eqs.~(\ref{eq34}) must be taken into
account).

Any solution to Eq.~(\ref{eq36}) determines a certain {\it form} of
NNM or its {\it spatial profile}. In our case, there are some
localized and delocalized modes among these solutions. In
particular, one of the solutions to Eq.~(\ref{eq36}) represents the
$\pi$-mode. Obviously, every localized NNM is an exact discrete
breather in accordance with its definition as spatially localized
and time-periodic vibration. The time dependence of the breather is
determined by the leading equation (\ref{eq35}) which can be solved
in terms of the Jacobi elliptic function
$\cn\left(\tau,\frac{1}{\sqrt{2}}\right)$ (see below). Note that
this result was obtained in \cite{l8}.

In this paper, we will be interested only in the breather solution
which is {\it symmetric} with respect to its center. Therefore, we
must assume the following relation to hold:
\begin{equation}\label{eq40}
k_{-n}=k_{n},\;\;\; n=-N..N.
\end{equation}
Taking into account Eq.~(\ref{eq40}) allows us to reduce by a half
the number of unknowns in Eq.~(\ref{eq36}). Let us write down these
equations for the cases $N=1$ and $N=2$ in the explicit form.

For N=1, we have the chain with three particles only
($\tilde{N}=3$). In this case, the dynamical equations (\ref{eq30})
read:
\begin{equation}
\begin{array}{l}
\ddot{x}_{-1}=-x_{-1}^3+\beta[(x_{0}-x_{-1})^3-(x_{-1}-x_1)^3],\\
\ddot{x}_0=-x_0^3+\beta[(x_{1}-x_0)^3-(x_0-x_{-1})^3],\\
\ddot{x}_1=-x_1^3+\beta[(x_{-1}-x_1)^3-(x_1-x_{0})^3].
\end{array}
\label{eq41}
\end{equation}
According to Eqs.~(\ref{eq32}) and (\ref{eq40}), the symmetric
breather pattern ($k_{-1}=k_1\equiv k$) reads
\begin{equation}\label{eq42}
\{kx_0(t),\; x_0(t),\; kx_0(t)\}.
\end{equation}
Substituting this pattern into Eqs. (\ref{eq41}), we obtain the
following algebraic equation from the condition of identity of all
these equations \footnote{We search for the solution with $k\neq1$
(the case $k=1$ corresponds to the delocalized mode).}:
\begin{equation}\label{eq43}
k(1+k)+\beta(1-k)^2(1+2k)=0.
\end{equation}
Since the root $k=k(\beta)$ of Eq.~(\ref{eq43}) is a function of
$\beta$, let us choose $\beta=0.3$ (below, it will be shown that the
breather is certainly stable for this value of $\beta$).

Using MAPLE, we obtain
\begin{equation}\label{eq44}
k=-0.29344944496399996095.
\end{equation}

Let us note that calculating all the Rosenberg modes with the aid of
Eq.~(\ref{eq36}) we used the MAPLE specification Digits=20.
Nevertheless, some values that have been calculated with such
precision will be, for compactness, presented with a smaller number
of digits.

Analogously, for the case $N=2$, the  chain consists of five
particles ($\tilde{N}=5$) and we must search the symmetric breather
pattern as follows
\begin{equation}\label{eq45}
\{k_2x_0(t),\; k_1x_0(t),\; x_0(t),\; k_1x_0(t)\;,k_2x_0(t)\}.
\end{equation}
Then the algebraic equations for $k_1$ and $k_2$ read
\begin{equation}
\begin{array}{lll}
k_1[-1+2\beta(k_1-1)^3)]=-k_1^3+\beta[(k_2-k_1)^3-(k_1-1)^3],\\
k_2[-1+2\beta(k_1-1)^3]=-k_2^3-\beta(k_2-k_1)^3
\end{array}
\label{eq46}
\end{equation}
and we obtain the following roots of these equations for
$\beta=0.3$:
\begin{equation}
\begin{array}{lll}
k_1=-0.29928831163054746768,\\
k_2=0.00359934143244973138.
\end{array}
\label{eq47}
\end{equation}

Continuing in this manner, we find symmetric breathers for the
chains with $\tilde{N}=7,9,11,13, etc$. particles. Some results of
these calculations are presented in Table \ref{t3}. Being calculated
with 20 digits, these results are practically the {\it exact}
breather solutions for the corresponding $K_4$ chains. Moreover,
comparing the profiles for the chains with $\tilde{N}=9$ and
$\tilde{N}=15$ particles, one can reveal that the further increase
of $\tilde{N}$ does not affect the spatial profile of the breather
solution. Indeed, the considered breathers demonstrate so strong
localization that the displacements of the particles  distant by
more than three lattice spacings from the breather center are
utterly insignificant (they don't exceed $10^{-20}$ and we denote
them in the tables by asterisk). Therefore, we conclude that the
profile for $\tilde{N}=19$ (and even for $\tilde{N}=5$) can be
considered as that for the infinite chain ($\tilde{N}=\infty$).

For the case {\it without} the on-site potential we have for
$\tilde{N}=3$: $k_0=1$, $k_{-1}=k_1=-0.5$. The similar results for
$\tilde{N}=5,9,15$ are presented in Table \ref{t3a}. From this
table, it is obvious that the results of Ref. \cite{l8} don't
correspond to the exact solution for the case of the infinite chain
since the author has used the profile only for $\tilde{N}=3$.
Indeed, the breathers considered in \cite{l8} are determined by the
pattern $\{0,..,0,0,-\frac{1}{2},1,-\frac{1}{2},0,0,...,0\}$, i.e.
all the particles outside of the central three-particle domain are
assumed to have zero amplitudes of oscillation. We cannot be sure
why the author of that paper refers to such  dynamical objects as
{\it exact} breathers despite they represent only a certain
approximation. On the other hand, it is evident from Tables
{\ref{t3}, \ref{t3a}} that dynamical objects with $\tilde{N}\geq 5$
particles can be practically considered as exact breathers.

Now let us continue to study the $K_4$-chain with on-site and
intersite potential for the case $\beta=0.3$.

\begin{table}
  \centering
  \caption{\label{t3} Spatial profiles of symmetric breathers in
  the $K_4$-chain with $\tilde{N}=5,9,15$ particles for
  $\beta=0.3$}
\begin{tabular}{|c|c|c|c|c|}
  \hline
    & $\tilde{N}=5$ & $\tilde{N}=9$ & $\tilde{N}=15$ \\
  \hline
$k_{-7}$ &  &  &  * \\
$k_{-6}$ &  &  &  *  \\
$k_{-5}$  &  &  & *   \\
$k_{-4}$  &  & * & *   \\
$k_{-3}$  &  & -0.6040174714525917849e-8 & -0.6040174714525917731994e-8  \\
$k_{-2}$ & 0.0035993414324497313812 & 0.0035993477082925520972 &  0.0035993477082925520972  \\
$k_{-1}$ & -0.29928831163054746768 & -0.29928831201300724704 & -0.2992883120130072470430  \\
$k_0$  & 1 & 1 & 1   \\
$k_1$ & -0.29928831163054746768 & -0.29928831201300724704 & -0.2992883120130072470419  \\
$k_2$ & 0.0035993414324497313812  & 0.0035993477082925520972 &  0.0035993477082925520972  \\
$k_{3}$  &  & -0.6040174714525917849e-8 & -0.6040174714525917731994e-8  \\
$k_{4}$  &  & * & *  \\
$k_5$  &  &  & *   \\
$k_6$  &  &  & *   \\
$k_7$  &  &  & *  \\
  \hline
\end{tabular}
\end{table}

\begin{table}
  \centering
  \caption{\label{t3a} Spatial profiles for symmetric breathers in
  the $K_4$-chain without on-site potential}
\begin{tabular}{|c|c|c|c|c|}
  \hline
    & $\tilde{N}=5$ & $\tilde{N}=9$ & $\tilde{N}=15$ \\
  \hline
$k_{-7}$ &  &  &  *  \\
$k_{-6}$ &  &  &  * \\
$k_{-5}$  &  &  & *   \\
$k_{-4}$  &  & * & *   \\
$k_{-3}$  &  & -0.17336102462887968846e-5 & -0.1733610246288796884739e-5  \\
$k_{-2}$ & 0.023048199202046015774 & 0.023050209905554654592 &  0.02305020990555465459272  \\
$k_{-1}$ & -0.52304819920204601577 & -0.52304847629530836653 & -0.5230484762953083665346  \\
$k_0$  & 1 & 1 & 1   \\
$k_1$ &  -0.52304819920204601577& -0.52304847629530836653 &  -0.5230484762953083665309  \\
$k_2$ & 0.023048199202046015774 & 0.023050209905554654592 &  0.02305020990555465459218  \\
$k_{3}$  &  & -0.17336102462887968846e-5 & -0.1733610246288796884198e-5  \\
$k_{4}$  &  & * & *   \\
$k_5$  &  &  & *  \\
$k_6$  &  &  & *   \\
$k_7$  &  &  & *  \\
  \hline
\end{tabular}
\end{table}

The time dependence of the breather solution is determined by the
leading equation (\ref{eq35}). For the symmetric breather
($k_{-1}=k_1\equiv k$), it can be written as follows:
\begin{equation}\label{eq50}
\ddot{x}_0+p^2x_0^3=0,
\end{equation}
where
\begin{equation}\label{eq51}
p^2=1+2\beta(1-k)^3.
\end{equation}
The parameter $p=p(\tilde{N})$ varies slightly  with changing the
number $\tilde{N}=2N+1$ of particles in the considered chain:
\begin{equation}
\begin{array}{lll}
p^2(3)= 2.2983734517955912888\\
p^2(5)= 2.3160362289746275590\\
p^2(7)= 2.3160362301367966972\\
p^2(9)= 2.3160362301367966972\;\;\; etc.
\end{array}
\label{eq52}
\end{equation}

For initial conditions
\begin{equation}\label{eq53}
x_0(0)=A_0, \;\;\;\; \dot{x}_0(0)=0
\end{equation}
the solution to Eq.~(\ref{eq50}) (see, for example, \cite{l8}) reads
\begin{equation}\label{eq54}
x_0(t)=A_0\cn\left(\omega t, \frac{1}{\sqrt{2}}\right),
\end{equation}
where the frequency $\omega$ is the linear function of the amplitude
$A_0$:
\begin{equation}\label{eq55}
\omega=p A_0.
\end{equation}

Here $\cn(\omega t, m)$ is the Jacobi elliptic function with the
modulus $m$ equal to $\frac{1}{\sqrt{2}}$. Note that such value of
the modulus is needed to eliminate the linear in $x_0(t)$ term,
because, in general case, the function $\cn(\tau,m)$ satisfies the
equation \footnote{This equation can be obtained using the
elementary formulas for the Jacobi elliptic functions (see, for
example, \cite{l16}).}:

$\cn''(\tau,m)+[1-2m^2]\cn(\tau,m)+2m^2\cn^3(\tau,m)=0.$

Introducing the new time and space variables $\tau$, $x(\tau)$
according to relations
\begin{equation}\label{eq56}
t=\frac{\tau}{pA_0}, \;\;\; x_0(t)=A_0x(\tau),
\end{equation}
we obtain from Eqs. (\ref{eq50}, \ref{eq53}) the following Cauchy
problem for the function $x(\tau)$ \footnote{We denote the
differentiation with respect to $t$ by dot, while that with respect
to $\tau$ by prime.}:
\begin{equation}\label{eq57}
x''+x^3(\tau)=0, \;\;\;x(0)=1, \;\;\;x'(0)=0
\end{equation}
with the solution
\begin{equation}\label{eq58}
x(\tau)=\cn\left(\tau,\frac{1}{\sqrt{2}}\right).
\end{equation}

As was the already mentioned, dynamical objects whose existence is
derived above demonstrate a strong localization and we can describe
them using the chain with $\tilde{N}=7$ particles only. Considering
longer chains would not contribute to the accuracy of description.

The strong localization occurs not only for $\beta=0.3$, but also
for others $\beta$ \footnote{Note that the breather loses its
stability for $\beta>0.554$.} (see Tables~\ref{t4},~\ref{t5}). As a
matter of fact, the localization varies, but this change is
completely negligible.

\begin{table}
  \centering
  \caption{\label{t4} Spatial profiles of symmetric breathers in
  the $K_4$-chain with $\tilde{N}=5,9,15$ particles for
  $\beta=0.5$}
\begin{tabular}{|c|c|c|c|c|}
  \hline
    & $\tilde{N}=5$ & $\tilde{N}=9$ & $\tilde{N}=15$ \\
  \hline
$k_{-7}$ &  &  &  *  \\
$k_{-6}$ &  &  &  *  \\
$k_{-5}$  &  &  & *  \\
$k_{-4}$  &  & * & *   \\
$k_{-3}$  &  & -0.83729113342838511470e-7 & -0.83729113342838510461e-7 \\
$k_{-2}$ & 0.0085070518871235750600 & 0.0085071418686266995085 & 0.0085071418686266994747 \\
$k_{-1}$ &  -0.38845832365012308590 & -0.38845833272969912287& -0.38845833272969912241  \\
$k_0$  & 1 & 1 & 1   \\
$k_1$ &  -0.38845832365012308590 &-0.38845833272969912287& -0.38845833272969912344 \\
$k_2$ & 0.0085070518871235750600 & 0.0085071418686266995085 & 0.0085071418686266995444  \\
$k_{3}$  &  & -0.83729113342838511470e-7 & -0.83729113342838512537e-7 \\
$k_{4}$  &  & * & *   \\
$k_5$  &  &  & *   \\
$k_6$  &  &  & *   \\
$k_7$  &  &  &  *  \\
  \hline
\end{tabular}
\end{table}

\begin{table}
  \centering
  \caption{\label{t5} Spatial profiles of symmetric breathers in
  the $K_4$-chain with $\tilde{N}=5,9,15$ particles for
  $\beta=0.6$}
\begin{tabular}{|c|c|c|c|c|}
  \hline
    & $\tilde{N}=5$ & $\tilde{N}=9$ & $\tilde{N}=15$ \\
  \hline
$k_{-7}$ &  &  &  *  \\
$k_{-6}$ &  &  &  *  \\
$k_{-5}$  &  &  & *   \\
$k_{-4}$  &  & * & *   \\
$k_{-3}$  &  & -0.15110644213594442410e-6 & -0.1511064421359444246249e-6  \\
$k_{-2}$ & 0.010331486554892818706 & 0.010331650738375748031 &  0.01033165073837574804346  \\
$k_{-1}$ & -0.41214169658344762796 & -0.41214171466223389870 & -0.4121417146622338988575  \\
$k_0$  & 1 & 1 & 1   \\
$k_1$ & -0.41214169658344762796 & -0.41214171466223389870 & -.04121417146622338985546 \\
$k_2$ & 0.010331486554892818706e-1 & 0.010331650738375748031 &  0.01033165073837574801942 \\
$k_{3}$  &  & -0.15110644213594442410e-6 & -0.1511064421359444235671e-6  \\
$k_{4}$  &  & * & *   \\
$k_5$  &  &  & *   \\
$k_6$  &  &  & *   \\
$k_7$  &  &  &  *  \\
  \hline
\end{tabular}
\end{table}

In conclusion, it is worth to emphasize that the spatial profile
$\{k_{-n},k_{-n+1},..,k_{-2},k_{-1},k_0=1,k_1,k_2,..,k_{n-1},k_n\}$
provided in Tables \ref{t3}, \ref{t3a}, \ref{t4}, \ref{t5} turns out
to be {\it universal} for the breathers with different amplitudes
$A_0$ (it does not depend on the amplitude), while the breather
frequency depends on $A_0$ linearly ($\omega=pA_0$).

\section{Stability of breathers in the $K_4$-chain \label{stab}}
\subsection{Linearization of the dynamical equations near the
breather solution}

To study the stability of a given periodic dynamical regime, in
accordance with the standard prescription of the linear stability
analysis, we must linearize the nonlinear equations of motion in the
vicinity of this regime (the breather solution, in our case) and
investigate the resulting linear equations with time-periodic
coefficients.

Let us start our stability analysis with the simplest example,
namely, we will consider the stability of the breather
\begin{equation}\label{eq60}
\vec{x}_b(t)=\{kx_0(t),\;x_0(t),\;kx_0(t)\}
\end{equation}
in the three-particle $K_4$-chain described by dynamical equations
(\ref{eq41}). To this end, we introduce an infinitesimal vector
\begin{equation}\label{eq61}
\vec{\delta}(t)=\{ \delta_{-1}(t),\;\delta_0(t),\;\delta_1(t)\},
\end{equation}
substitute the vector $\vec{x}(t)=\vec{{x}}_b(t)+\vec{\delta}(t)$
into Eqs.~(\ref{eq41}) and linearize these equations with respect to
$\delta_j\;\;(j=-1,\;0,\;1)$.

As the result of this procedure, we obtain the linearized system
\begin{equation}\label{eq62}
\ddot{\vec{\delta}}(t)=-3x_0^2(t)A\,\vec{\delta}(t)
\end{equation}
with the symmetric matrix
\begin{equation}\label{eq63}
A=   \left(%
\begin{array}{ccc}
  \mu+\nu & -\mu & 0 \\
  -\mu & 1+2\mu & -\mu \\
  0 & -\mu & \mu+\nu \\
\end{array}
\right),
\end{equation}
where $\mu=\beta(1-k)^2,\;\;\; \nu=k^2.$ Here the coefficient $k$ is
determined by the algebraic equation (\ref{eq43}), while $x_0(t)$ is
the solution to the leading equation (\ref{eq50}) with the initial
conditions (\ref{eq53}) (for $\beta=0.3$ \,
$p^2=p^2(3)=2.2983734517955912888$).

It can be easily shown that Eq.~(\ref{eq62}) is valid for an
arbitrary value $\tilde{N}$, but the corresponding matrix $A$, in
this case, turns out to be more complicated. For example, for the
$K_4$-chain with five particles ($\tilde{N}=5$) we have
\begin{equation}\label{eq64}
A=\left(
\begin{array}{ccccc}
  \eta+k_2^2 & -\eta & 0 & 0 & 0 \\
  -\eta & \eta+\mu+k_1^2 & -\mu & 0 & 0 \\
  0 & -\mu & +1+2\mu & -\mu & 0 \\
  0 & 0 & -\mu & \eta+\mu+k_1^2 & -\eta \\
  0 & 0 & 0 & -\eta & \eta+k_2^2 \\
\end{array}
\right),
\end{equation}
where $\mu=\beta(1-k_1)^2$,  $\eta=\beta(k_1-k_2)^2$. Here $k_1$ and
$k_2$ are determined by Eqs.~(\ref{eq46}) [for $\beta=0.3$, their
numerical values are given by Eqs.~(\ref{eq47})], while $x_0(t)$ is
the solution to Eq.~(\ref{eq50}) with
$p^2=p^2(5)=2.3160362289746275590$.

 The specific structure of the linearized system (\ref{eq62})
allows us to make an essential step in the simplification of our
further stability analysis. Indeed, let us pass from the vector
variable $\vec{\delta}(t)$ to a new variable
$\tilde{\vec{\delta}}(t)$ whose definition involves a {\it
time-independent} orthogonal matrix $S$:
\begin{equation}\label{eq69}
\vec{\delta}(t)=S\,\tilde{\vec{\delta}}(t).
\end{equation}
Substituting $\vec{\delta}$ in such form into Eq.~(\ref{eq62}) and
multiplying this equation by the matrix $\tilde{S}$ from the left
($\tilde{S}=S^{-1}$ is the transpose of $S$), we obtain
\footnote{The tildes in $\tilde{\delta}$ and $\tilde{S}$ are used in
different sense: $\tilde{\vec{\delta}}$ is the new vector variable
with respect to the old variable $\vec{\delta}$, while $\tilde{S}$
is the transpose of $S$.}
\begin{equation}\label{eq70}
\ddot{\tilde{\vec{\delta}}}=-3x_0^2(t)(\tilde{S}AS)\tilde{\vec{\delta}}.
\end{equation}

On the other hand, the matrix $A$ is symmetric \footnote{This
property is a consequence of the fact that the linearized system
(\ref{eq62}) can be written in the form
$\ddot{\vec{\delta}}=J(t)\,\vec{\delta} $via the Jacobi matrix
$J(t)$ which is constructed from the second partial derivatives of
the total potential energy of the considered chain.} and, therefore,
there exists an orthogonal matrix $S$ transforming the matrix $A$ to
the fully diagonal form $A_{diag}$:
\begin{equation}\label{eq71}
\tilde{S}AS=A_{diag}.
\end{equation}
If we find such matrix $S$, the linearized system (\ref{eq70})
decomposes into $\tilde{N}$ {\it independent} differential equations
\begin{equation}\label{eq72}
\ddot{\tilde{\delta}}_j+3x_0^2(t)\lambda_j\tilde{\delta}_j=0,\;\;\;j=-N..N,
\end{equation}
where $\lambda_j$ are the eigenvalues of the matrix $A$. Moreover,
solving the eigenproblem $A$$\vec{y}$=$\lambda$$\vec{y}$ for the
matrix $A$, we obtain not only $\lambda_j$ for Eq.~(\ref{eq72}), but
also the explicit form of the matrix $S$ from Eq.~(\ref{eq69}): its
columns turn out to be the eigenvectors $\vec{y}_j$  ($j=-N..N$) of
the matrix $A$.

Each equation of (\ref{eq72}) represents the linear differential
equation with {\it time-periodic} coefficient. The most well-known
differential equation of this type is the Mathieu equation
\begin{equation}\label{eq74}
\ddot{z}+[a-2q\cos(2t)]z=0.
\end{equation}
The ($a-q$) plane for this equation splits into regions of stable
and unstable motion \cite{l16}. If parameters ($a$,$q$) fall into a
stable region, $z(t)$ that is small at the initial instant $t=0$
continues to be small for all times $t>0$ (the case of Lyapunov
stability). In the opposite case, if $z(0)$ is a small value (even
infinitesimal), $z(t)$ will begin to grow rapidly for $t>0$
(Lyapunov instability). Actually, we must analyze the stability of
the {\it zero} solution of Eq. (\ref{eq74}). In the next subsection,
we study the analogous stability properties of the equations
(\ref{eq72}).

\subsection{Investigation of the basic equation}

Let us consider Eq.~(\ref{eq72}) in more detail. The time-periodic
function $x_0(t)$, entering this equation, is the solution to the
Cauchy problem [see Eq.~(\ref{eq50}, \ref{eq53})]
\begin{center}
$\ddot{x_0}+p^2 x_0^3(t)=0, \;\;\;\;
x_0(0)=A_0,\;\;\;\;\dot{x}_0(0)=0.$
\end{center}

On the other hand, the change of variables (\ref{eq56}) allows us to
eliminate the dependence of the Cauchy problem on the breather
amplitude $A_0$ (see Eq.~(\ref{eq57})). Using the same change of
variables in Eq.~(\ref{eq72}), we get the equation
\begin{equation}\label{eq80}
\ {z}''_j+\frac{3\lambda_j}{p^2}x_0^2(\tau) z_j(\tau)=0,
\end{equation}
where $z_j(\tau)=\tilde{\delta}_j(\dfrac{\tau}{pA_0})$. For the sake
of clarity of the further investigation, it is convenient to drop
the subscript $j$ from this equation and rewrite it in the form
\begin{equation}\label{eq81}
\ {z}''+\Lambda x_0^2(\tau)z(\tau)=0,
\end{equation}
where $\Lambda=\dfrac{3\lambda}{p^2}$.

Our stability analysis of the breathers in the $K_4$-chain is based
on Eq.~(\ref{eq81}) and we will refer to it as "basic equation". One
important conclusion can be deduced immediately from this equation,
namely, the stability of our breathers does not depend on their
amplitudes $A_0$! The fact is that the amplitude $A_0$ is not
contained in Eq.~(\ref{eq81}) neither explicitly, nor implicitly
($p$ is expressed via $k_1$, $k_{-1}$ which, in turn, are determined
by an algebraic equation independent of $A_0$).

\begin{figure}
\includegraphics[width=155mm,height=85mm]{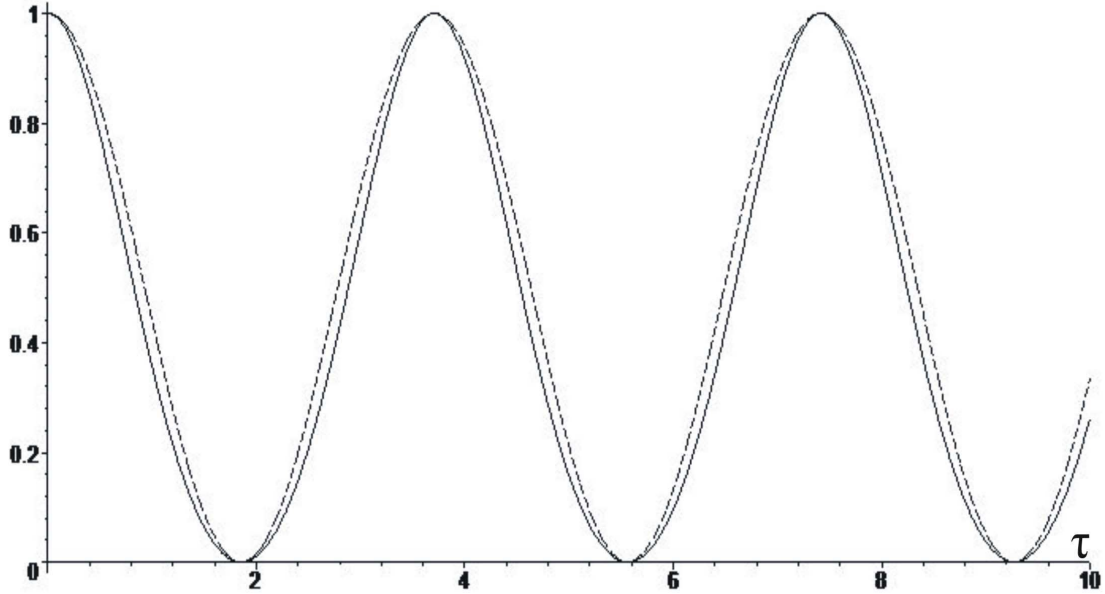}
\caption{\label{fig2} Comparison of the functions
$\cn^2\left(\tau,\frac{1}{\sqrt{2}}\right)$ and
$\frac{1}{2}[1+\cos(\Omega \tau)]$.}
\end{figure}

It is interesting that our basic equation (\ref{eq81}) can be
reduced to the Mathieu equation (\ref{eq74}) within a certain
approximation. Indeed, taking into account Eq.~(\ref{eq58}), we can
write
\begin{equation}\label{eq82}
x_0^2(\tau)=\cn^2\left(\tau,\frac{1}{\sqrt{2}}\right)\approx
\frac{1}{2}(1+\cos(\Omega\,\tau)),
\end{equation}
where $\Omega\approx1.6944$. Actually, we approximate the periodic
function $\cn^2\left(\tau,\frac{1}{\sqrt{2}}\right)$ by the two
first terms of its Fourier series. Surprisingly, this fit turns out
to be a very good approximation, as one can see from
Fig.~\ref{fig2}, where the function $\cn^2(\tau,\frac{1}{\sqrt{2}})$
and $\frac{1}{2}(1+\cos(\Omega \tau))$ are shown by the solid and
dashed lines, respectively. Note that several first terms of the
Fourier series for
$\cn^2\left(\tau,\frac{1}{\sqrt{2}}\right)$ read\\
$\cn^2\left(\tau,\frac{1}{\sqrt{2}}\right)=0.4569+0.4972\cos(\Omega
\tau)+0.0429\cos(2\Omega \tau)+0.0028 \cos(3\Omega \tau)+... $.

\begin{figure}
\includegraphics[width=105mm,height=85mm]{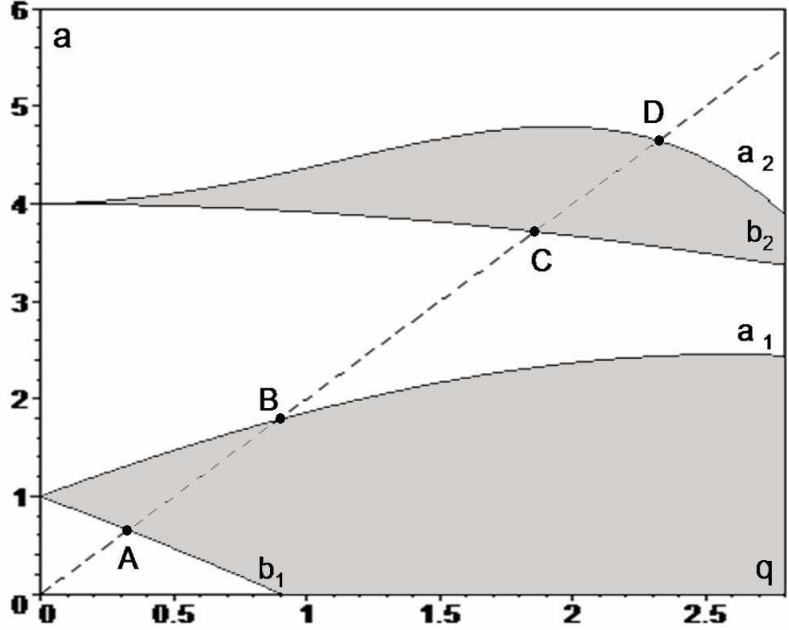}
\caption{\label{fig3} The regions of stable and unstable motion for
the Mathieu equation.}
\end{figure}

Introducing a new time variable according to the formula
$\Omega\tau=2\tilde{t}+\pi$, we obtain from Eq.~(\ref{eq81}):
\begin{equation}\label{eq83}
\ddot{\tilde{z}}+\frac{2\Lambda}{\Omega^2}[1-\cos(2\tilde{t})]\cdot
\tilde{z}(\tilde{t})=0.
\end{equation}
Obviously, this is the Mathieu equation (\ref{eq74}) with
\begin{equation}\label{eq84}
a=\frac{2\Lambda}{\Omega^2}\;\;\; { \mbox{and}} \;\;\;
q=\frac{\Lambda}{\Omega^2}.
\end{equation}
Thus, we obtain a certain relation between two arbitrary parameters
of Eq.~(\ref{eq74}):
\begin{equation}\label{eq85}
a=2\,q.
\end{equation}
Let us now recall the well-known stability diagram for the Mathieu
equation (see, for example,~\cite{l16}). In Fig.~\ref{fig3}, the
regions of unstable motion are shaded.

The boundary lines $a_i(q)$, $b_i(q)$ ($i=1,2$) of these regions can
be approximated by the following series:
\begin{center}
$b_1(q)=1-q-\dfrac{q^2}{8}+\dfrac{q^3}{64}-\dfrac{q^4}{1536}+...$,\medskip\\
$a_1(q)=1+q-\dfrac{q^2}{8}-\dfrac{q^3}{64}-\dfrac{q^4}{1536}+...$,\medskip\\
$b_2(q)=4-\dfrac{q^2}{12}+\dfrac{5q^4}{13824}+...$,\medskip\\
$a_2(q)=4+\dfrac{5q^2}{12}-\dfrac{763q^4}{13824}+...$.\\
\end{center}

In Fig.~\ref{fig3}, we also depict the straight line $a=2q$
according to Eq.~(\ref{eq85}) and four points ($A$,$B  $,$C$,$D$) of
the intersection of this line with the boundary curves $b_i(q)$,
$a_i(q)$ ($i=1,2$). The following values of $q$ correspond to these
points of intersection:
\begin{center}
$q_A=0.329$,    $\;\;q_B=0.890$,    $\;\;q_C=1.858$, $\;\;q_D=2.321.
$
\end{center}

On the other hand, $q=\frac{\Lambda}{\Omega^2}$ (see
Eq.~(\ref{eq84})) and, therefore, we can approximately find the
values of the parameter $\Lambda$ corresponding to the boundaries of
stable and unstable motion for the basic equation (\ref{eq81}). In
this way, we obtain the regions represented in the first column of
Table \ref{t6}.

Thus, there exist regions of stable and unstable motion for our
basic equation (\ref{eq81}) in accordance with the numeric values of
the parameter $\Lambda$ entering this equation.

The above approach based on the Mathieu equation analysis not only
sheds light on the main properties of Eq.~(\ref{eq81}), but allows us
to arrive at some approximate results presented in Table \ref{t6}.
On the other hand, we can obtain analogous results with high
precision analyzing the basic equation (\ref{eq81}) with the aid of
the Floquet method.

To this end, we construct the $2\times 2$ monodromy matrix by
integrating Eq.~(\ref{eq81}) twice (with the initial conditions
$z(0)=1$, $z'(0)=0$ and $z(0)=0$, $z'(0)=1$) over one period of the
function $\cn\left(\tau, \frac{1}{\sqrt{2}}\right)$ and calculate
its multiplicators. If the absolute value of a multiplicator exceeds
unity by more then $10^{-5}$, we identify the case of unstable
motion.

The results obtained by the Floquet method prove to be surprising!
Indeed, all boundary values of $\Lambda$ are {\it integer numbers},
at least up to $10^{-5}$:
\begin{center}
$\Lambda_A=1$, $\;\;\Lambda_B=3$, $\;\;\Lambda_C=6$,
$\;\;\Lambda_D=10$.
\end{center}
We suspect that this property of Eq.~(\ref{eq81}) can be proved
exactly, but we could not find a proof of this conjecture
\footnote{However, we can point out to one argument concerning the
plausibility of this conjecture. According to the Floquet theory,
the solutions corresponded to the boundaries of the regions of
stable and unstably motion must be strictly periodic functions. On
the other hand, one can obtain a periodic solution to the basic
equation (\ref{eq81}) for $\Lambda=1$:
$z(\tau)=x(\tau)=\cn\left(\tau,\frac{1}{\sqrt{2}}\right)$. Indeed,
in this case, Eq.~(\ref{eq81}) reduces to Eq.~(\ref{eq57}) whose
solution is the Jacobi elliptic function
$\cn\left(\tau,\frac{1}{\sqrt{2}}\right)$.}.

Thus, for Eq.~(\ref{eq81}), we obtain the regions of stable and
unstable motion presented in the second column of Table \ref{t6}.

\begin{table}
  \centering
  \caption{\label{t6} Regions of stable and unstable motion for
  Eq.~(\ref{eq81})}
  \begin{tabular}{|c|c|c|}
    \hline
    Analysis based on the Mathieu equation& Exact Floquet analysis & Stable or unstable motion \\
    \hline
    $0<\Lambda<0.945$ & $0<\Lambda<1$ & stable \\
    $0.945<\Lambda<2.554$ & $1<\Lambda<3$ & unstable \\
    $2.554<\Lambda<5.335$ & $3<\Lambda<6$ & stable \\
    $5.335<\Lambda<6.665 $ & $6<\Lambda<10$ & unstable \\
    \hline
  \end{tabular}
\end{table}

\subsection{Quasibreathers in the $K_4$-chains \label{qk4c}}

The above results for the basic equation (\ref{eq81}) allow us to
make a final step in our breather stability analysis. Indeed, we
have reduced this analysis  to the problem of stability of the {\it
zero solutions} to $\tilde{N}$ individual equations (\ref{eq80}). A
given breather will be stable, if {\it all} the values
$\Lambda_j=\dfrac{3\lambda_j}{p^2}\;\;\;(j=-N..N)$ from
Eq.~(\ref{eq80}) fall into certain regions of stability motion of
the basic equation (\ref{eq81}).

With high precision, we have computed  the eigenvalues $\lambda_j$
of the  matrix $A$ [see, for example, Eqs.~(\ref{eq62}-\ref{eq64})]
and the values $\Lambda_j=\dfrac{3\lambda_j}{p^2}$ entering the
basic equation~(\ref{eq81}) for the  $K_4$-chain with  different
number of particles ($\tilde{N}$) and different values of the
parameter $\beta$ determining the strength of the intersite
potential. It has been revealed that  all $\Lambda_j$, with
exception of $\Lambda_1$, depend considerably on $\beta$ and
slightly on $\tilde{N}$.

On the other hand, $\Lambda_1=3$ for all $\tilde{N}$ and $\beta$ and
this {\it constant} coincides, at least up to $10^{-10}$, with the
boundary between the first region of unstable and the second region
of stable motion of the basic equation~(\ref{eq81}) [see
Table~\ref{t6}]. Because of the importance of the equality
$\Lambda_1=3$, we prove it analytically for $\tilde{N}=3$ in
Appendix 1.

According to the Floquet  theory, this means that a strictly
time-periodic solution corresponds to $\Lambda_1=3$. Moreover, the
eigenvector $\vec{V}_1$ of the matrix $A$, corresponding to
$\Lambda_1$, remarkably coincides with high precision (see also
Appendix 1 for the analytical proof of this fact  for the case
$\tilde{N}=3$) with the {\it spatial profile} of the considered
breather for all $\tilde{N}$ and $\beta$. Therefore, the
infinitesimal perturbation {\it along} the vector $\vec{V}_1$ does
not relate to  the stability of  the breather: it leads only to the
infinitesimal increasing of the breather's amplitude. Thus, studying
the breather stability, we must consider only $\Lambda_j$ with
$j\neq1$.

In Table~\ref{t7}, for $\beta=0.3$, we present the eigenvalues
$\lambda_j$ of the matrix $A$ and the corresponding values
$\Lambda_j$ for the $K_4$-chains with $\tilde{N}=3,5,7,9$ particles.
It may be concluded, from this table, that for $\tilde{N}\geq7$ only
$\Lambda_2$, $\Lambda_3$, $\Lambda_4$ and $\Lambda_5$ are of
considerable magnitude and correct values of these $\Lambda_j$, at
least up to $10^{-5}$, can be found already from the case
$\tilde{N}=5$. From Table~\ref{t6}, we also see that all $\Lambda_j$
\ ($j>1$) fall into the first region of stability ($0<\Lambda_j<1$)
and, therefore, the considered breathers are {\it stable} for
$\beta=0.3$.

\begin{table}
  \centering
  \caption{\label{t7} Eigenvalues of the matrix A of the linearized
  dynamical system for the $K_4$-chain with different number of
  particles $\tilde{N}$ ($\beta=0.3$)}
\begin{tabular}{|c|c|c|c|}
  \hline
  \multicolumn{2}{|c|}{$\tilde{N}=3$} & \multicolumn{2}{|c|}{$\tilde{N}=5$} \\
  \hline
  $\lambda_1=2.298373452$& $\Lambda_1=3$ & $\lambda_1=2.316036234$& $\Lambda_1=3$ \\
  $\lambda_2=0.5880160166$& $\Lambda_2=0.7675201995$ &  $\lambda_2=0.6248090298$& $\Lambda_2=0.8093254611$  \\
  $\lambda_3=0.2934494449$& $\Lambda_3=0.3830310231$ &  $\lambda_3=0.3226215801$& $\Lambda_3=0.4178970630$  \\
   &  & $\lambda_4=0.02626701223$& $\Lambda_4=0.03402409500$   \\
   &  & $\lambda_5=0.02530830363$& $\Lambda_5=0.03278226390$ \\
  \hline
  \multicolumn{2}{|c|}{$\tilde{N}=7$} & \multicolumn{2}{|c|}{$\tilde{N}=9$}\\
  \hline
  $\lambda_1=2.316036234$& $\Lambda_1=3$  &  $\lambda_1=2.316036234$& $\Lambda_1=3$   \\
  $\lambda_2=0.6248090398$& $\Lambda_2=0.8093254740$  &  $\lambda_2=0.6248090398$& $\Lambda_2=0.8093254740$  \\
  $\lambda_3=0.3226216094$& $\Lambda_3=0.4178971017$  &  $\lambda_3=0.3226216094$& $\Lambda_3=0.4178971017$ \\
  $\lambda_4=0.02627089232$& $\Lambda_4=0.03402912093$  &   $\lambda_4=0.02627089232$& $\Lambda_4=0.03402912093$ \\
  $\lambda_5=0.02531216336$ & $\Lambda_5=0.03278726346$ &  $\lambda_5=0.02531216336$ & $\Lambda_5=0.03278726346$ \\
  $\lambda_6=0.3886030305e-5$& $\Lambda_6=0.5033639268e-5$   &  $\lambda_6=0.3886030305e-5$& $\Lambda_6=0.5033639268e-5$   \\
  $\lambda_7=0.3886011392e-5$& $\Lambda_7=0.5033614770e-5$   &  $\lambda_7=0.3886011392e-5$& $\Lambda_7=0.5033614770e-5$ \\
   &  &  $\lambda_8=0.1094511318e-16$& $\Lambda_8=0.1417738596e-16$   \\
   &  &  $\lambda_9=0.1094511318e-16$& $\Lambda_9=0.1417738596e-16$  \\
  \hline
\end{tabular}
\end{table}

\begin{figure}
\includegraphics[width=155mm,height=85mm]{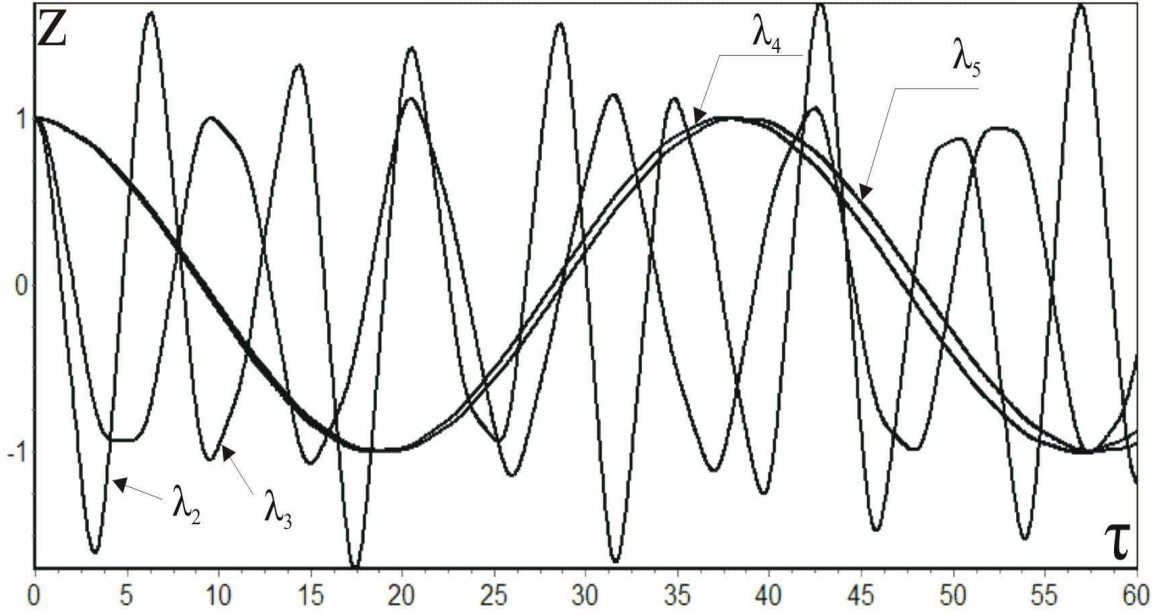}
\caption{\label{fig4} Solutions to the basic equation (\ref{eq81})
corresponding to different values of $\Lambda$.}
\end{figure}

In Fig.~\ref{fig4}, we present the solution to Eq.~(\ref{eq81}) for
$\Lambda_2$, $\Lambda_3$, $\Lambda_4$, $\Lambda_5$ corresponding to
a certain initial amplitude (the value of this amplitude is
inessential because Eq.~(\ref{eq81}) is linear). They {\it are not
periodic}, but {\it stationary} solutions in the sense that their
amplitudes don't increase infinitely in time, as it occurs for the
case where $\Lambda_j$ fall into an unstable region. Thus, if we are
in a close (even infinitesimal) vicinity of a given breather, i.e.
if all $\tilde{\delta}_j(t)$ [see Eq.~(\ref{eq72})] are small values
at the initial instant $t=0$, they continue to be small for all
later times $t>0$. Then according to Eq.~(\ref{eq69}), the smallness
of the Chebyshev norm $\|\tilde{\delta}_j(t)\|_c$ implies the
smallness of $\|\delta_j(t)\|_c$.

On the other  hand, it follows from Eq.~(\ref{eq69}) that the
solution to the original nonlinear equation (\ref{eq30}) {\it is not
periodic} in any small vicinity of the exact breather! Indeed,
because of the relation $\vec{\delta}(t)=S\tilde{\vec{\delta}}(t)$,
each $\delta_j(t)$ is a certain linear combination of all the
$\tilde{\delta}_j(t) \;\;\;(j=-N..N)$, but individual
$\tilde{\delta}_j(t)$ are, in general, quasiperiodic functions.
[Even if certain $\tilde{\delta}_j(t)$ would be periodic, their
periods are independent of each other and, therefore, the total
solution $\delta_j(t) \;\;\;(j=-N..N)$ will not be periodic in any
case.]. In other words, we arrive at the conclusion that arbitrary
small vicinity of the exact breather solution consist of {\it
quasiperiodic} solutions which can be naturally called {\it
quasibreathers}.

Moreover, in the case of the $K_4$-chains with $\beta=0.3$, these
quasibreathers turn out to be {\it stable} dynamical objects.
Indeed, in a sufficiently small vicinity of the exact breather, the
quasibreathers are described by the vector
$\vec{x}(t)=\vec{x}_b(t)+\vec{\delta}(t)$ [see Eqs.~(\ref{eq60}) and
(\ref{eq61})] where dynamics of $\vec{\delta}(t)$ is determined by
the linearized system (\ref{eq62}). Above we have demonstrated that
all $\parallel\delta_j(t)\parallel_c$ are certainly small for any
$t>0$. Therefore, one can conclude that not only the considered
breather solution, but also the quasibreather solutions, which are
close to it, must be stable in the Chebyshev norm.

Actually, in the $K_4$-space, there exist a certain one-dimensional
family of the exact breathers with different amplitudes $A_0=x_0(0)$
and with the same spatial profile (see,
Tables~\ref{t3},~\ref{t3a},~\ref{t4},~\ref{t5}). A straight line
corresponds to this family in the $\tilde{N}$-dimensional space of
all the conceivable initial conditions $x_i(0) \;\;\;(i=-N..N).$ It
is practically impossible to tune exactly onto this specific line in
the many-dimensional space.

On the other hand, in any vicinity of this line, we have to deal
with the quasibreathers: the different particles possess different
frequencies and, moreover, these frequencies evolve in time. Such a
behavior of the individual particles in a quasibreather vibration in
the $K_4$-chain can be illustrated by the method used in Sec. 2 for
studying the James breathers.

We investigate the stability of the quasibreather solutions by
direct numerical integration of the differential
equations~(\ref{eq30}) of the considered chain over large time
intervals. To this end, we choose a certain initial deviation
\begin{equation}\label{eq800}
\vec{\delta}(0)=\varepsilon\cdot\{\delta_{-N}(0),...,\delta_0(0),...,\delta_N(0)\}
\end{equation}
from the exact spatial profile $\vec{x}_b(0)$ of a given breather,
where $\varepsilon$ is a small parameter, while $\delta_n(0)$ \
[$n=-N..N$] are {\it random} numbers whose absolute values don't
exceed unity. Then we solve Eq.~(\ref{eq30}) with initial condition
$\vec{x}(0)=\vec{x}_b(0)+\vec{\delta}(0)$, \ $\dot{\vec{x}}(0)=0$
and examine the numerical solution
$\vec{x}(t)=\{x_{-N}(t),...,x_0(t),...,x_N(t)\}$ after a long time.
We can scan any vicinity of the exact breather by varying
$\varepsilon$ and the random sequence $\{\delta_n(0) \ | \
n=-N..N\}$ from Eq.~\ref{eq800}. In Table~\ref{t9}, we present the
results of such a calculation for the $K_4$-chain with $\tilde{N}=7$
and $\beta=0.3$. We have used the fourth-order Runge-Kutta method
with time step $h=0.001$ and integrated Eq.~(\ref{eq30}) up to
$t=1500T_0$, where $T_0$ is the period of the  $\pi$-mode
($T_0=\pi$). The frequencies ($\omega_j$) of only five breather
particles ($j=-2..2$) have been taken into account, because the
vibrational amplitudes of the particles with $j=\pm3$ are very small
(they are of the  order of $10^{-8}$).

\begin{table}
  \centering
  \caption{Derviations in frequencies of different breather (quasibreather) particles}\label{t9}
\begin{tabular}{|c|c|c|c|c|c|c|}
  \hline
    & $\varepsilon=0$ & $\varepsilon=10^{-7}$ & $\varepsilon=10^{-6}$ & $\varepsilon=10^{-5}$ & $\varepsilon=10^{-4}$ & $\varepsilon=10^{-3}$ \\
  \hline
  $\omega_{-2}$ & 1.289333684 & 1.289324677 & 1.289226737 & 1.288414794 & 1.252673857 &   \\
  $\omega_{-1}$ & 1.289333684 & 1.289333939 & 1.28933311 & 1.289297098 & 1.289077907 & 1.287039954 \\
  $\omega_0$ & 1.289333684 & 1.28933432 & 1.289333134 & 1.289332824  & 1.2893981453 & 1.2933772905  \\
  $\omega_1$ & 1.289333684 & 1.289333847 & 1.289332133 & 1.289281901  &  1.2890993617 &  1.2887949577 \\
  $\omega_2$ & 1.289333684 & 1.289320432 & 1.289203072 & 1.288248186 &  1.2767350998 & \\
  \hline
  $\eta$ & 0 & 2.8923591607e-6 & 2.911839882e-5 & 2.397956354e-4 & 7.103135734e-3 &  1.8891448851e{-3} \\
  \hline
  $\eta_{max}$ & 1.261367077e{-10} & 1.0051826702e-6 & 1.0272244997e-5 & 1.002013005e-4 & 1.000467033e{-2} &   2.443460953e-1 \\
  \hline
\end{tabular}
\end{table}

\begin{figure}[h!t!b!]
\includegraphics[width=100mm,height=50mm]{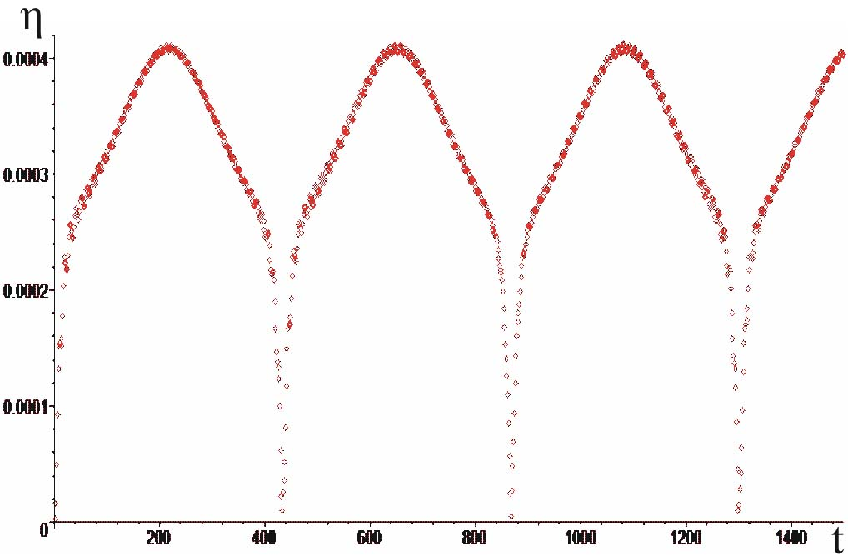}
\caption{\label{fig9} Mean square deviations $\eta=\eta(t_k)$ in
frequencies of the breather (quasibreather) particles for
$\varepsilon\approx10^{-9}$. Time $t$ is given in periods $T_0$.}
\end{figure}

\begin{figure}[h!t!b!]
\includegraphics[width=100mm,height=55mm]{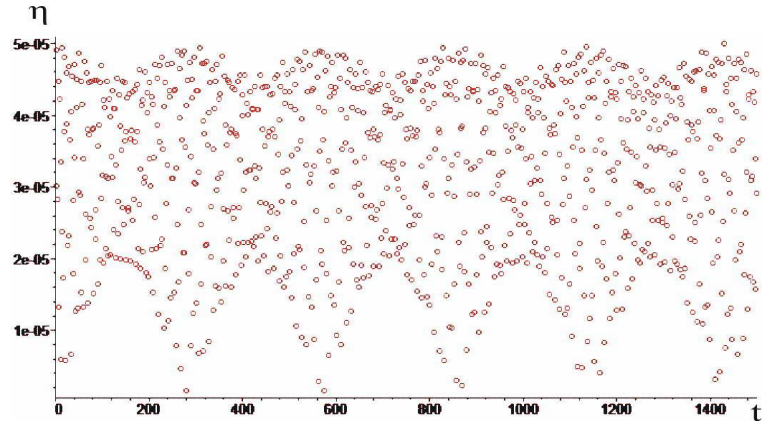}
\caption{\label{fig10} Mean square deviations $\eta=\eta(t_k)$ in
frequencies of the breather (quasibreather) particles for
$\varepsilon\approx10^{-6}$. Time $t$ is given in periods $T_0$.}
\end{figure}

In Table~\ref{t9},  we also present the mean square deviations
$\eta=\eta(t_k)$ for $\omega_j$ determined by Eq.~(\ref{eq11}),
which have been computed for $t_k\approx1500T_0$, and the maximal
values of $\eta(t_k)$ for the interval $0<t_k<1500T_0$. The fact is
that $\eta(t_k)$ varies on the considered time interval (see, for
example, Figs.~\ref{fig9},~\ref{fig10}) and, therefore, $\eta_{max}$
is a more relevant characteristics of the deviations in frequencies
of the individual particles.

From Table~\ref{t9}, one can see the specific quasibreather
phenomenon, namely, the deviations in the frequencies $\omega_j$ of
the individual particles increases with increasing the parameter
$\varepsilon$, characterizing the deviation of the quasibreather
shape from that of the exact breather in the Chebishev norm. It is
important to emphasize that despite different  particles possess
slightly different frequencies $\omega_j$, the quasibreathers are
{\it stable} dynamical objects. Indeed, we did not observe any decay
of these objects for $\varepsilon\leq10^{-2}$ up  to $t=10^6T_0$
($\eta_{max}$, even for such time interval, does not practically
differ from those presented in Table~\ref{t9}).

However,  for $\varepsilon>10^{-2}$, we have observed the decay of
the quasibreathers which manifest itself in appearance of
appreciable vibrational amplitudes of those particles whose
amplitudes were practically equal to zero in the exact breather
solution.

In conclusion, let us return to Figs.~\ref{fig9} and ~\ref{fig10},
where we depict $\eta(t_k)$ as a function of the subsequent instants
$t_k$ for which the frequencies $\omega_j$ were calculated.
Sometimes, the function $\eta(t_k)$ demonstrate regular
oscillations, sometimes, practically chaotic oscillations. Such a
behavior can be understood, if one takes into account that the
displacement of each particle is a superposition of different
quasiperiodic functions, as it was shown above in the present
section.

\subsection{Stability of breathers with respect to strength of
intersite potential}

We consider the $K_4$-chain determined by the potential energy
(\ref{eq8a}) inducing the Newton dynamical equations
($\ddot{x}_n=-\dfrac{\partial U}{\partial{x_n}}$) in the
form~(\ref{eq30}). Let us discuss the stability of the breathers and
quasibreathers in this chain with respect to the parameter $\beta$
which determines the strength of the intersite part of the potential
energy relative to its on-site part. Eigenvalues $\lambda_j$ of the
matrix $A$ from Eq.~(\ref{eq62}) and, therefore, the corresponding
values $\Lambda_j$ of the parameter $\Lambda$ entering
Eq.~(\ref{eq81}), depend on $\beta$: $\lambda_j=\lambda_j(\beta)$,
$\Lambda_j=\Lambda_j(\beta)$.

\begin{figure}
\includegraphics[width=160mm,height=165mm]{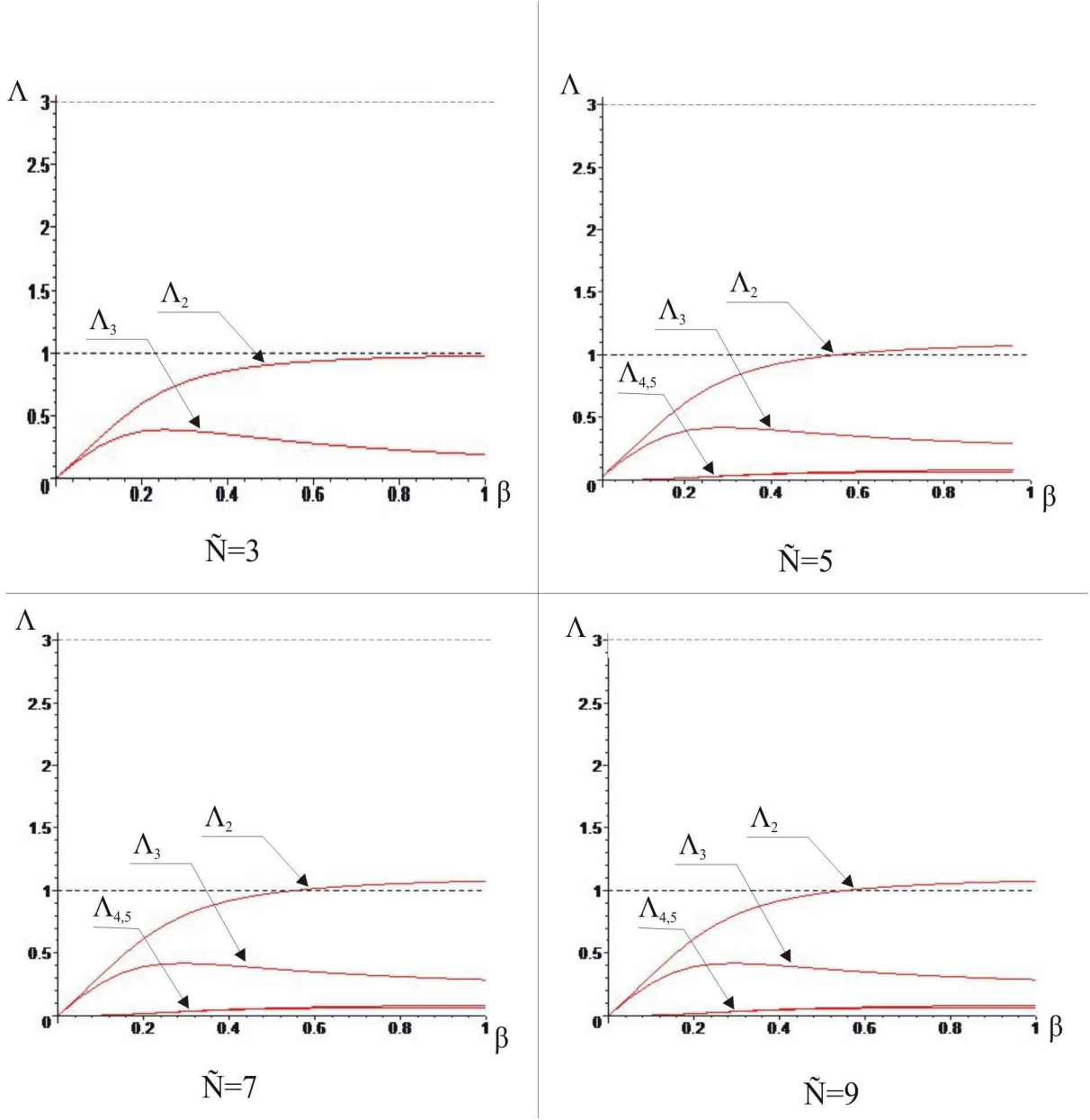}
\caption{\label{fig8} The functions $\Lambda_j(\beta)$ for different
eigenvalues of the matrix A for the $K_4$-chain with
$\tilde{N}=3,\,5,\,7,\,9$ particles. Note that the plots of
$\Lambda_4(\beta)$ and $\Lambda_5(\beta)$ practically coincide with
each  other (see Table \ref{t7} for $\tilde{N}=9$).}
\end{figure}

In Fig.~\ref{fig8}, we present $\Lambda_j(\beta)$ with $j=2,3,4,5$
for the $K_4$-chain with $\tilde{N}=3,\,5,\,7,\,9$ particles. These
$\Lambda_j(\beta)$ are of the more significant values, as it follows
from Table~\ref{t7}. All $\Lambda_j(\beta)$ which are not depicted
in Fig.~\ref{fig8} are small {\it positive} numbers. Then one can
conclude that for $\beta\in[0,0.554]$ all $\Lambda_j(\beta)$ remain
in the first stability region ($0<\Lambda<1$) of the basic
equation~(\ref{eq81}). As it has been already discussed this
demonstrates the stability of the considered breathers (and,
therefore, quasibreathers which are close to them) when $\beta$
increases from zero up to $\beta_c=0.554$. (Note that our breathers
are unstable for $\beta<0$).

On the other hand, we find from Fig.~\ref{fig8} that $\Lambda_2$
intersects the upper boundary ($\Lambda=1$) of the first stability
region for $\tilde{N}>3$ when $\beta$ exceeds $0.554$. This implies
the loss of stability of the breathers in the considered
$K_4$-chain.

Thus, the intersite part of the potential with $\beta>0$ {\it must
not be too large} with respect to its on-site part ($\beta<0.554$)
for breathers (quasibreathers) to be stable.

Finally, let us point out an one-dimensional subspace of the space
of all possible displacements which becomes unstable when $\beta$
intersects the critical value $\beta_c=0.554$. This subspace is
determined by the eigenvector $\vec{V}$ corresponding to
$\lambda_2$:

\begin{center}
$\vec{V}=[0.05266431305, -0.7051428715, 0, 0.7051428725,
-0.05266431306].$
\end{center}

\section{Conclusion}

The main conclusion of the present paper is that the conventional
view on the discrete breathers as {\it strictly time periodic} and
spatially localized  dynamical objects must be revised in a certain
sense. Indeed, it has been shown here that for the James breathers
\cite{l4,l5} in the $K_2-K_3-K_4$ chain as well as for the breathers
in the $K_4$-chain with on-site and intersite potentials we actually
deal with the dynamical objects representing {\it quasibreathers}:
there are certain deviations in the vibrational frequencies of the
individual particles. These deviations can be characterized by the
mean square deviation which certainly exceeds the possible numerical
errors. Moreover, for the case of the $K_4$-chain, we have performed
a {\it rigorous} investigation of the {\it existence} and {\it
stability} of such quasibreathers. For the $K_4$-chain, the exact
breathers exist only along a certain line in the many-dimensional
space of all the possible initial conditions, and it is actually
impossible to tune precisely onto this line in any physical and
computational experiments.

In some of our numerical experiments with the $K_4$-chain, the
deviations in frequencies of the vibrating particles from the
average quasibreather frequency $\omega_b$ attained 1\%, but
possibly these deviations can considerably exceed this value for
more realistic models.

The deviations in frequencies of the individual particles (and in
frequency of the given particle over time) result in some change of
the breather Fourier spectrum, namely, instead of the ideal lines at
the breather frequency $\omega_b$ and its multiples there appear
certain (possibly, narrow) packets of the Fourier components near
the ideal breather lines and near zero frequency. This effect is
difficult to reveal with the aid of the numerical Fourier analysis
and we prefer to study the deviations in vibrational frequencies of
the individual particles in the straightforward way.

It is essential, that the above described deviations in frequencies
of the individual vibrating particles, in general, don't mean an
onset of the breather decay. We have demonstrated this fact with the
example of stable quasibreathers in the $K_4$-chain. For this case,
we succeeded in proving that these quasibreathers turn out to be
stable up to a {\it certain strength} of the intersite potential
with respect to the on-site potential.

We conjecture that the results obtained in the present paper for two
particular cases (the James breathers and the breathers in the
$K_4$-chain) are also valid for the general case and, therefore, one
must speak about quasibreathers rather than about strictly
time-periodic breathers.

Finally, let us note that the term "quasiperiodic breathers" is used
in literature for dynamical objects different from the
quasibreathers considered in the present paper (for example, see
\cite{l17} and the references therein). Indeed, the former possess
several basis frequencies (and their integer linear combinations) in
the Fourier expansion with substantial amplitudes, while the latter
possess only one basis frequency  $\omega_b$ (and its multiples), as
well as many small components with frequencies different from
$n\omega_b$. The quasiperiodic breathers exist only in rather
specific cases, while quasibreathers seam to be the generic
dynamical objects.

\begin{acknowledgments}
We are very grateful to Prof.~V.~P.~Sakhnenko for his friendly
support and to O.~E.~Evnin for his valuable help with the language
corrections in the text of this paper.
\end{acknowledgments}

\appendix
\section{ }

In Sec.~\ref{qk4c}, using straightforward numerical calculations, we
have demonstrated that the largest eigenvalue $\lambda_1$ of the
matrix $A$ [see Eqs.~(\ref{eq62}-\ref{eq64})] corresponds to the
boundary between the first unstable and the second stable regions
for the basic equation (\ref{eq81}). Moreover, the eigenvector
$\vec{V_1}$, associated with $\lambda_1$, determines the direction
of the infinitesimal perturbation along the considered breather.

Below we prove these propositions analytically for the $K_4$-chain
with $\tilde{N}=3$ particles. Let us consider Eq.~(\ref{eq62})
\begin{equation}\label{A1}
    \ddot{\vec\delta}=-3x_0^2(t)\cdot A\vec{\delta}
\end{equation}
with matrix $A$ determined by Eq.~(\ref{eq63}). The parameter $k$
entering the matrix $A$ determines the spatial profile
\begin{equation}\label{A2}
    \{k, 1, k\}
\end{equation}
of the breather
\begin{equation}\label{A3}
    \{kx_0(t), x_0(t), kx_0(t)\},
\end{equation}
while time dependence of the breather particles can be obtained from
equations (\ref{eq50},\ref{eq51}):
\begin{equation}\label{A4}
\ddot{x}_0+p^2x_0^3=0,
\end{equation}
\begin{equation}\label{A5}
p^2=1+2\beta(1-k)^3.
\end{equation}

On the other hand, parameter $k=k(\beta)$ is a function of the
intersite potential stregth $\beta$, as it follows from
Eqs.~(\ref{eq43}):
\begin{equation}\label{A6}
    k(1+k)+\beta(1-k)^2(1+2k)=0.
\end{equation}

The eigenvalues $\lambda_j$ ($j=1, 2, 3$) of the matrix $A$ can be
calculated from the characteristic equation of this matrix in the
following form:
\begin{equation}\label{A7}
    \lambda_{1,3}(\beta)=\frac{1}{2}[(1+k^2)+3\beta(1-k)^2]\pm\frac{1}{2}(1-k)\sqrt{(1+k)^2+2\beta(1-k^2)+9\beta^2(1-k)^2},
\end{equation}
\begin{equation}\label{A8}
 \lambda_2(\beta)=k^2+\beta(k-1)^2.
\end{equation}

In principle, one can obtain these eigenvalues as explicit functions
of the intersite potential strength $\beta$ from Eq.~(\ref{A6}) and
substitute it into Eqs.~(\ref{A7}) and (\ref{A8}). This way is too
cumbersome and we prefer to use the following trick. Let us express
$\beta$ via $k$ from Eq.~(\ref{A6}) and substitute $\beta=\beta(k)$
into Eqs.~(\ref{A7}, \ref{A8}). Then, the square root entering
Eq.~(\ref{A7}) can be explicitly extracted and written in the form
\begin{equation}\label{A9}
    \frac{(1+k)(1+2k^2)}{1+2k}.
\end{equation}
As a consequence of this extraction, we obtain
\begin{equation}\label{A10}
    \lambda_1=\frac{1+2k^3}{1+2k},
\end{equation}
\begin{equation}\label{A11}
    \lambda_3=-k.
\end{equation}
The same substitution of $\beta=\beta(k)$ into Eqs.(\ref{A8}) and
(\ref{A5}) permits us to write $\lambda_2$ and $p^2$ as follows:
\begin{equation}\label{A12}
     \lambda_2=\frac{k(1-2k^2)}{1+2k},
\end{equation}
\begin{equation}\label{A13}
     p^2=\frac{1+2k^3}{1+2k},
\end{equation}
Comparing Eqs.~(\ref{A10}) and (\ref{A13}), we obtain
\begin{equation}\label{A14}
    \lambda_1(\beta)=p^2(\beta).
\end{equation}
This is an important result since our basic equation~(\ref{eq81})
reads
\begin{equation}\label{A15}
   z''+\Lambda x_0^2(\tau)x(\tau)=0,
\end{equation}
with
\begin{equation}\label{A16}
    \Lambda=\dfrac{3\lambda}{p^2(\beta)}.
\end{equation}
Then, we can conclude that
$\Lambda_1=\dfrac{3\lambda_1}{p^2(\beta)}=3$ and, therefore,
$\Lambda_1$ is a constant with respect to the intersite potential
strength $\beta$! Moreover, as already has been discussed in
Sec~\ref{qk4c}, this value turns out to be the exact boundary
between the first region of unstable motion ($1<\Lambda<3$) and the
second region of stable motion ($3\leq\Lambda\leq6$) of the basic
equation~(\ref{A15}).

Let us recall  that {\it numerical} calculations have convinced us
with high precision that $\Lambda_1=3$ not only for the case
$\tilde{N}=3$, but for any other number of the particles in the
$K_4$-chain. Moreover, it can be proved that $\Lambda_1=m$ for the
uniform potential of the {\it arbitrary order} $m$ (in the present
paper, we consider only the case $m=3$).

The eigenvectors $\vec{V}_j$ ($j=1,2,3$) of the matrix $A$
corresponding to $\lambda_j$ from Eqs.~(\ref{A10}-\ref{A12}) can be
written after the substitution $\beta$ from Eq.~(\ref{A6}) as
follows:
\begin{equation}\label{eqA30}
    \vec{V}_1=[k,1,k],
    \vec{V}_2=[-1,0,1],
    \vec{V}_3=[1,-2k,1],
\end{equation}
where $k=k(\beta)$. We see that $V_1=[k,1,k]$ represents the vector
which coincides exactly with the spacial profile (\ref{A2}) of the
considered breather. On the other hand, $\vec{V}_j$ ($j=1,2,3$) are
the eigenvectors of the matrix $A$ [see Eq.~(\ref{A1})] of the
dynamical equations of the $K_4$-chain linearized near the
breather~(\ref{A3}). Therefore, $\vec{V}_1=[k,1,k]$, corresponding
to $\Lambda=3$, determines the direction {\it along} our breather in
the three-dimensional space of all the displacements of the
particles. In Section~\ref{stab}, using numeric calculations with
high precision, we have already arrived at the conclusion that this
result turns out to be correct not only for the case $\tilde{N}=3$,
but also for the $K_4$-chain with arbitrary number of particles.

\end{document}